\documentclass{proposalnsf}
\usepackage{times}
\usepackage{epsfig}
\usepackage{graphicx}
\usepackage{amsfonts}       
\usepackage{amsmath}
\usepackage{amssymb}
\usepackage{amsthm}
\usepackage{commath}
\usepackage{bm}
\usepackage{url}
\newcommand{\comment}[1]{}

\title{\Large \bf Sampling for View Synthesis: From Local Light Field Fusion to Neural Radiance Fields and Beyond}
\author{Ravi Ramamoorthi \\ University of California San Diego\footnote{Email: ravir@cs.ucsd.edu.  The author is also currently affiliated wtih NVIDIA.}}
\date{}

\begin{document}
\maketitle

\vspace*{-0.2in}
\section*{Abstract}
Capturing and rendering novel views of complex real-world scenes is a long-standing problem in computer graphics and vision, with applications in augmented and virtual reality, immersive experiences and 3D photography.  The advent of deep learning has enabled revolutionary advances in this area, classically known as image-based rendering.  However, previous approaches require intractably dense view sampling or provide little or no guidance for how users should sample views of a scene to reliably render high-quality novel views.  Local light field fusion proposes an algorithm for practical view synthesis from an irregular grid of sampled views that first expands each sampled view into a local light field via a multiplane image scene representation, then renders novel views by blending adjacent local light fields.  Crucially, we extend traditional  plenoptic sampling theory to derive a bound that specifies precisely how densely users should sample views of a given scene when using our algorithm.  We achieve the perceptual quality of Nyquist rate view sampling while using up to 4000x fewer views.  Subsequent developments have led to new scene representations for deep learning with view synthesis, notably neural radiance fields, but the problem of sparse view synthesis from a small number of images has only grown in importance.  We reprise some of the recent results on sparse and even single image view synthesis, while posing the question of whether prescriptive sampling guidelines are feasible for the new generation of image-based rendering algorithms.

\section{Introduction and Basics of Light Field Sampling Theory}
This article is written in response to the Frontiers of Science Award generously granted in 2024 to the paper~\cite{llff} on {\em Local Light Field Fusion: Practical View Synthesis with Prescriptive Sampling Guidelines} from SIGGRAPH 2019 (published in the ACM Transactions on Graphics).  The joint first-authors of this work were then UC Berkeley Ph.D. students Ben Mildenhall and Pratul Srinivasan\footnote{The following year, Ben, Pratul, Matt Tancik, Jon Barron, Ren and I published subsequent work on NeRF: Rpresenting Scenes as Neural Radiance Fields for View Synthesis at the 2020 European Conference on Computer Vision, which has now broadly been adopted in the field, and was recognized with an inaugural Frontiers of Science Award in 2023.  For their dissertations, including subsequent recognition with back-to-back Frontiers of Science Awarded papers Local Light Field Fusion and NeRFs, Pratul and Ben received an Honorable Mention for the 2021 ACM Doctoral Dissertation Award (awarded in 2022).}, collaborators at FYusion, Rodrigo Ortiz-Cayon and Abhishek Kar, Ren Ng at UC Berkeley, former UCSD Postdoc (current TAMU Faculty) Nima Kalantari and myself.  All authors made key contributions to enable the groundbreaking algorithm and results from local light field fusion, and many have continued to push the field forward in a remarkable sequence of subsequent papers.  

The local light field fusion paper (LLFF)\footnote{Since this article explains the LLFF paper, significant language is taken directly from that article, and not explictly put in quotes.} tackles the core view synthesis problem within image-based rendering (IBR).  To create compelling virtual experiences, we need to immerse the viewer within the scene.  That is, from a few images of the scene, we need to be able to synthesize new views to allow a user to walk around the scene, view it from different directions, zoom in or out, and change their viewpoint or orientation.  This can be seen within the context of sampling and reconstructing/interpolating the light field of the scene.\footnote{The light field is one of the core concepts in image-based rendering, and predates modern computing approaches, traced back to Gershun's work~\cite{Gershun} and even earlier attempts to build what today we would consider light field cameras~\cite{Ives,Lippman}.  We are inspired by the plenoptic function work of Adelson et al.~\cite{Adelson} and the light field and lumigraph papers from SIGGRAPH 96~\cite{Gortler,Levoy}.}

LLFF seeks to develop a simple sample-and-reconstruct approach to view synthesis.  As with any sampling problem, one is ultimately limited by the Nyquist rate, depending on the frequencies in the original signal.  Ideally, one simply captures images on a (semi)-regular grid, and interpolates them.  However, the Nyquist rate view sampling is intractable for scenes at interactive distances as the required sampling rate increases linearly with the reciprocal (disparity) of the nearest scene depth.  For a scene with a subject at a depth of 0.5 meters captured by a mobile phone camera with a 64$^\circ$ field of view and rendered at 1 megapixel resolution, the required sampling rate is an intractable 2.5 million images per square meter.  LLFF seeks to employ sophisticated light field sampling analysis and a data structure based on a multiplane image~\cite{TinghuiMag} to reduce the Nyquist rate by a factor of about 4000$\times$, thus enabling a tractable number of images (typically 20-50) to be used for view synthesis on fairly large baselines, with casual mobile phone capture and only simple light field reconstruction and interpolation.  

\begin{figure}
\includegraphics[height=2.0in]{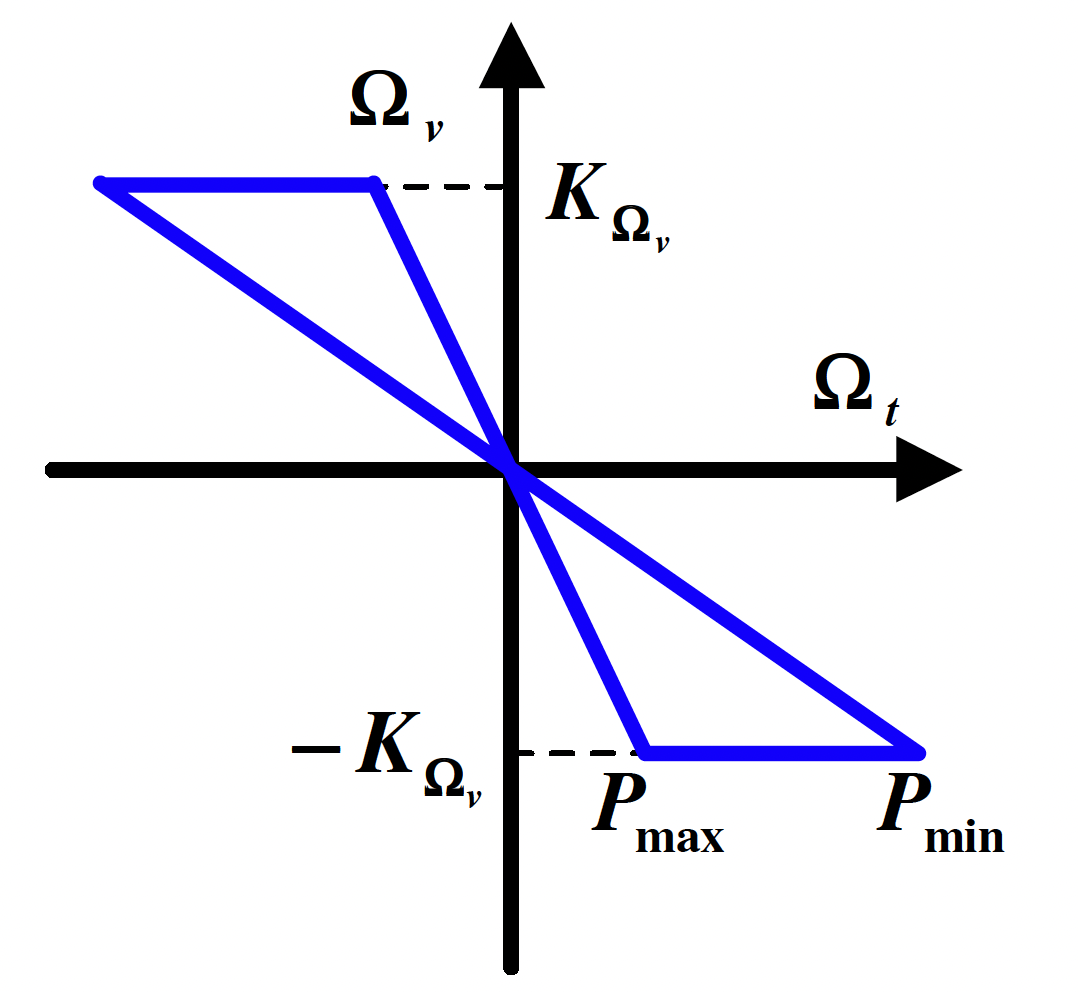} 
\includegraphics[height=2.0in]{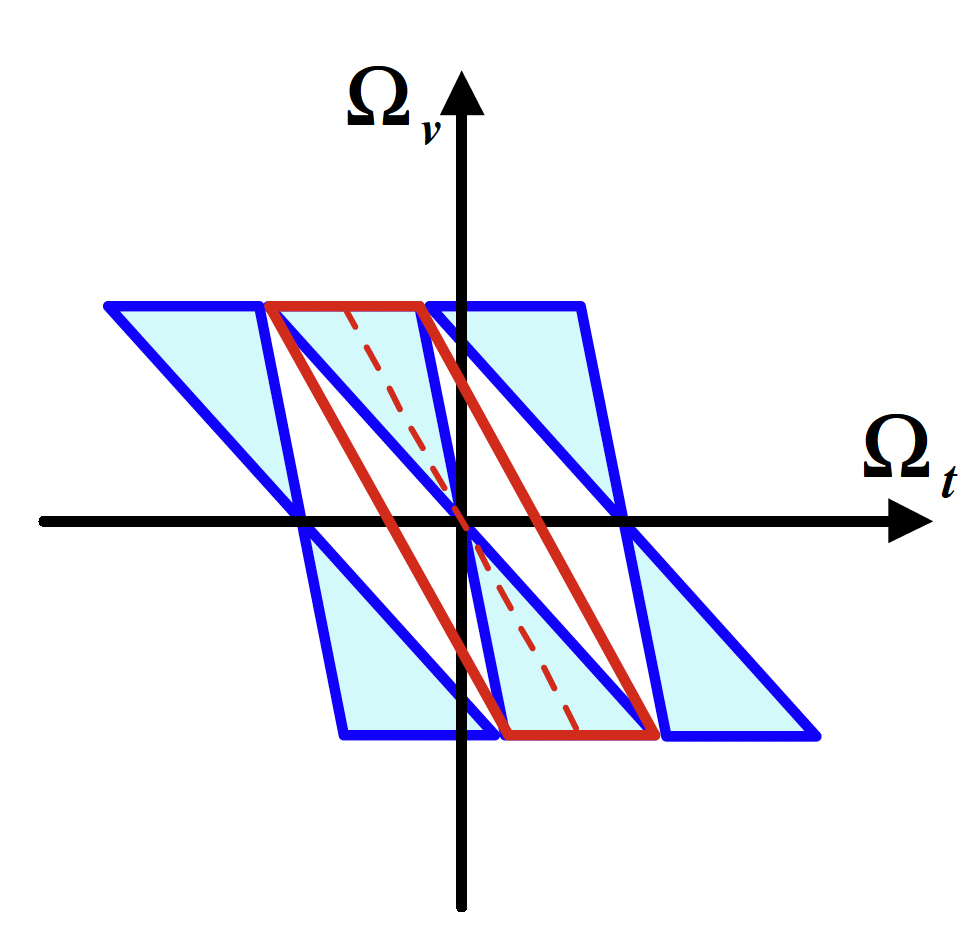} 
\includegraphics[height=2.0in]{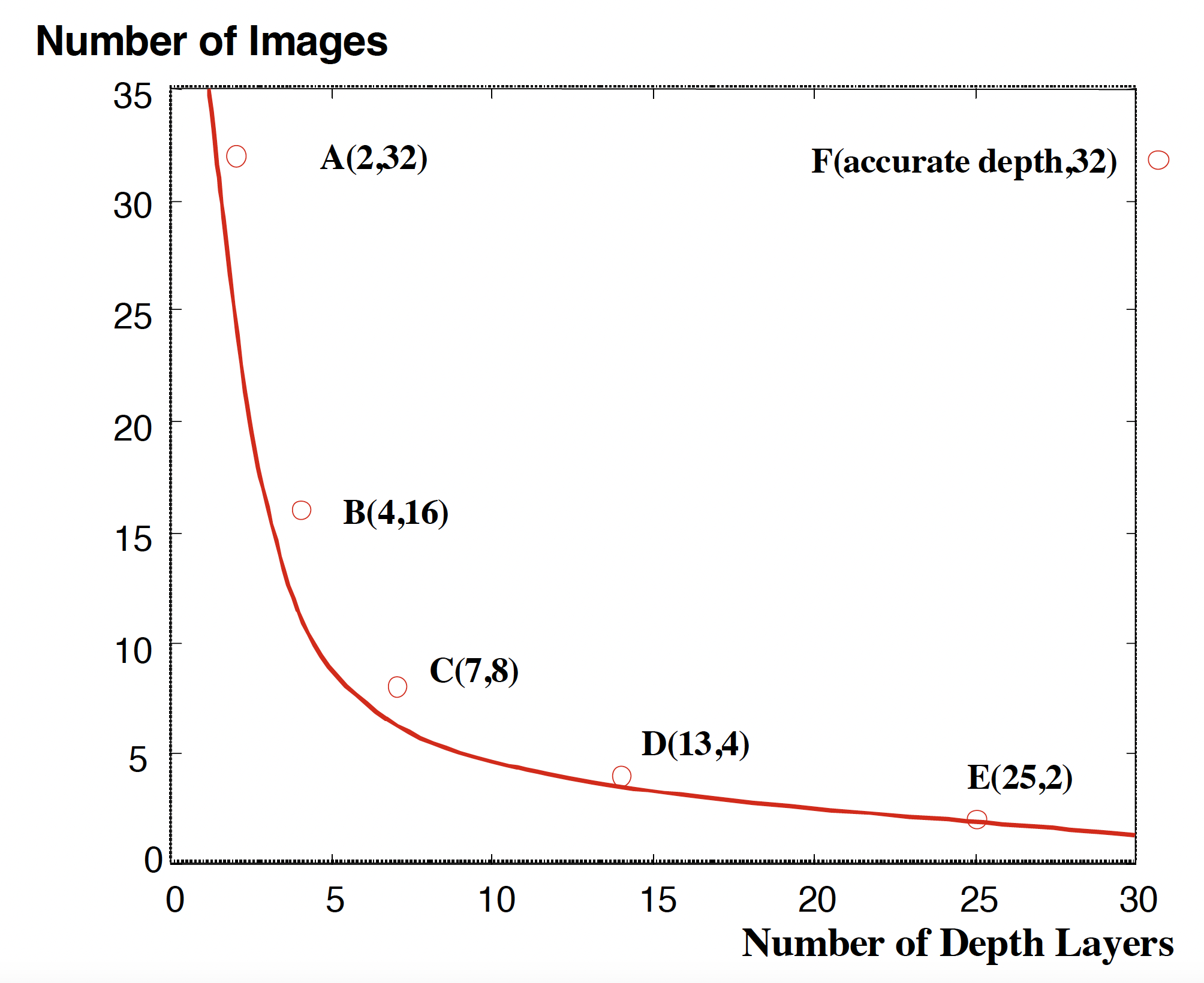}
\caption{\em \small Some basic results from the plenoptic sampling paper~\cite{plenoptic}.  On the left is the basic form of the double wedge spectrum.  In the middle we show packing of replicas with just sparse enough sampling so the central double wedge can be isolated with a parallelogram reconstruction filter.  On the right is the geometry-image sampling curve showing how fewer images are needed with more depth layers.  Figures taken from Chai et al.  Local light field fusion extends these results to account for occlusions, and shows how to apply the method for prescriptive view sampling guidelines with rigorous bounds in the context of modern deep learning multiplane image prediction methods.  }
\label{fig:plenopticsampling}
\end{figure}

The LLFF work leverages and builds strongly on a seminal paper from SIGGRAPH 2000 on plenoptic sampling~\cite{plenoptic}.  That paper addresses the question of the minimum sampling rate curve for Nyquist rate sampling in image-based rendering.  Crucially it argues for a joint geometry-image sampling rate (see Fig.~\ref{fig:plenopticsampling}), where the number of views depends on the accuracy with which the scene geometry is known.  Note that the early light field rendering paper~\cite{Levoy} argued that no intermediate representation was required, and one could simply interpolate the originally captured rays.  However, the concurrent lumigraph paper~\cite{Gortler} did introduce an approximate geometric model, and plenoptic sampling~\cite{plenoptic} argues that the accuracy of the geometric model influences the sampling rate needed for image capture.  Intuitively, if the surface is Lambertian, and we know the geometry exactly, then only one image needs to be captured as all other views will see the same color.    On the other hand, if we know nothing about the scene geometry, then many more images would potentially be needed to ensure that interpolation can be done without any aliasing artifacts.  

Plenoptic sampling~\cite{plenoptic} makes many groundbreaking contributions.  First, it performs a theoretical analysis of the Fourier spectrum for three-dimensional scenes.  The result is shown in Fig.~\ref{fig:plenopticsampling} (left).  As can be seen, the frequency is a classial double wedge spectrum, with the slopes bounded by the minimum and maximum depths of objects in the scene.  This double-wedge spectrum is fundamental not just in the local light field fusion paper, but we have also applied it to many problems in Monte Carlo rendering and other applications, briefly discussed later in this article.  Based on this analysis, a geometry-image sampling rate curve can be derived, shown on the right of Fig~\ref{fig:plenopticsampling}, which indicates how the number of views can potentially be reduced dramatically from $32\times 32$ to only $2\times 2$ or smaller, when the geometry can be accurately localized.  

In the interests of keeping this article simpler and more readable, we will not go into the mathematical details of plenoptic sampling (though we highly recommend authors read the original articles).  Instead, as noted at the end of the plenoptic sampling work~\cite{plenoptic}, we can give an intuitive characterization in terms of disparity.  The disparity is proportional to inverse depth, and essentially characterizes how the pixel position of a particular point moves when the camera shifts locations (translates).  In the limit that the point is infinitely distant, the disparity is zero, as the direction to that object remains the same as the camera translates (as is the case, for example to celestial bodies such as the sun or moon).  On the other hand, the image sampling rate can be limited by the largest disparity (closest point).  With no knowledge of geometry, the camera sampling rate must be set so that the disparity of any part of the scene for adjacent cameras is less than one pixel.   In essence, this is the Nyquist sampling rate, determined by the closest scene point or maximum disparity.  

The key insight is in using geometry to reduce the image sampling rate.  The disparity argument above is relevant if we are directly interpolating, effectively assuming an infinite depth.  However, if we use the actual depth of scene points to perform the interpolation, then one image suffices as noted above.  In general, if we can isolate the scene into a band of depths, then we may place the plane for reprojecting samples at some intermediate or mean (technically harmonic mean) depth, and the ``disparity'' is now only with respect to this mean depth.  As such, the closer we can isolate depth ``layers'' in the scene, the fewer images we need by leveraging the geometry in the scene.  From the perspective of formal frequency analysis, one effectively has a tight parallelogram filter that enables isolation of the central replica, despite other replicas caused by sampling (see middle of Fig.~\ref{fig:plenopticsampling}). 

In LLFF, we extend the previous sampling analysis to directly specify how users should sample input images for reliable high-quality view synthesis with antialiased rendering using (at the time) modern deep learning methods.  At the same time, this is not really a paper on deep learning; indeed, learning is used only to create a multi-plane image~\cite{Shade,SzeliskiGolland,TinghuiMag}.  Rather, LLFF is a rare paper in the modern deep learning world that aims to provide formal guarantees on the result, developing a sampling theory and prescriptive view synthesis guidelines.  The current article does not aim to be a comprehensive survey of image-based rendering, either prior to the LLFF work, or on subsequent developments, or indeed to even provide an in-depth review of the LLFF paper itself.\footnote{In terms of early work on image-based rendering, we encourage readers to look at volume 2 of the seminal graphics papers brought out for the 50th SIGGRAPH conference~\cite{seminal2}; in chronlogical order these papers are~\cite{ChenWilliams,McMillan,Levoy,Gortler,Deb1,environmentmatting,Debevec1,Wood,ulr,NayarGI,phototourism}.}
Rather, we provide some perspectives on the key results, discuss similar developments in other areas like Monte Carlo rendering, and briefly discuss new developments involving radiance fields.  
Finally, we return to the challenge of low-sample view synthesis and a theoretical framework for this work.   

\section{Local Light Field Fusion}
\begin{figure}
\begin{centering}
\includegraphics[width=0.75\columnwidth]{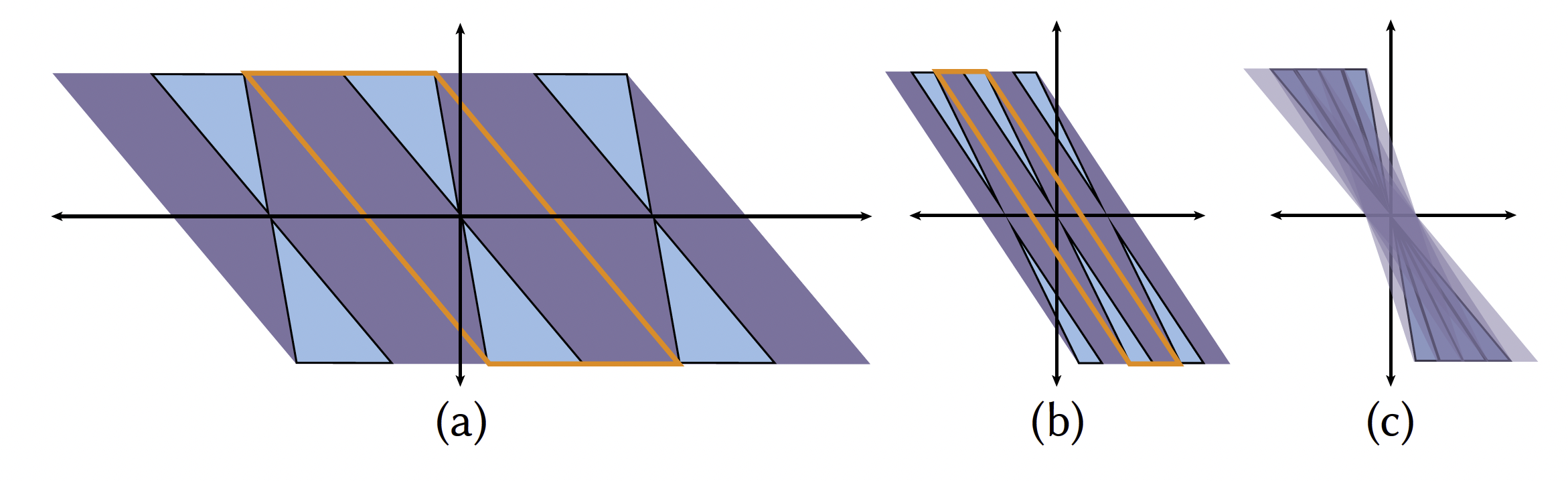}
\caption{\small \em Local light field fusion extends traditional plenoptic sampling to consider occlusions when reconstructing a continuous light field from MPIs. (a) Considering occlusions expands the Fourier support to a parallelogram (the Fourier support without occlusions is shown in blue and occlusions expand the Fourier support to additionally include the purple region) and doubles the Nyquist view sampling rate. (b) As in the no-occlusions case, separately reconstructing the light field for D layers decreases the Nyquist rate by a factor of D. (c) With occlusions, the full light field spectrum cannot be reconstructed by summing the individual layer spectra because the union of their supports is smaller than the support of the full light field spectrum (a). Instead, we compute the full light field by alpha compositing the individual light field layers from back to front in the primal domain.}
\label{fig:occlusions}
\end{centering}
\end{figure}

The first contribution of local light field fusion is in giving precise conditions for practical Nyquist rate view sampling.  The theory of plenoptic sampling can be extended, based on the work of Zhang and Chen~\cite{zhangchen} to account properly for occlusions.  As shown in Fig.~\ref{fig:occlusions}, one can consider the frequency spectrum as being a convolution with the occluder spectrum (multiplication by the occlusion mask in the primal domain).  This extends the double wedge to a parallelogram, which can only be packed half as densely as the original double wedge.  It is possible to derive precisely that the required maximum camera sampling interval $\bigtriangleup_u$ for a light field with occlusions is:
\begin{equation}
\bigtriangleup_u \leq \frac{1}{2 K_x f \left(1/z_{\min} - 1/z_{\max}\right)},
\end{equation}
where $f$ is the focal length, and $z_{\min}$ and $z_{\max}$ are minimum and maximum depths.  $K_x$ is the highest spatial frequency in the sampled light field, determined by the highest spatial frequency in the continuous light field $B_x$ and the camera spatial resolution $\bigtriangleup_x$ as , 
\begin{equation}
K_x = \min\left(B_x,\frac{1}{2\bigtriangleup_x}\right).
\end{equation}

One can now break the scene into layers using a multi-plane image.  Following~\cite{TinghuiMag}, this is a set of fronto-parallel RGB$\alpha$ planes, evenly sampled in disparity within a reference camera's view frustum.  Each image is ``promoted'' to an MPI through a simple deep-learning algorithm that looks at the image and its neighbors.  Rendering from an MPI is straightforward, just involving image compositing.  Key for our purposes is that plenoptic sampling theory shows that decomposing a scene into $D$ depth ranges and separately sampling the light field within each range allows the camera sampling interval to be increased by a factor of $D$. This is because the spectrum of the light field emitted by scene content within each depth range lies within a tighter double-wedge that can be packed D times more tightly than the full scene's double-wedge spectrum.  A key aspect of the local light field fusion paper is to extend this simple analysis, conducted without considering occlusions, to also handle occlusions, taking advantage of the predicted opacities in a multiplane image.  We show that with this extended analysis, we can still increase the require camera sampling interval by a factor of $D$ when there are $D$ depth layers so that, 
\begin{equation}
\bigtriangleup_u \leq \frac{D}{2 K_x f \left(1/z_{\min} - 1/z_{\max}\right)}.
\end{equation}
A further condition is obtained from the finite field of view, requiring that each point in the scene's bounding volume should fall within the frusta of at least two neighboring sampled views.  It can be shown that this can be expressed in terms of the width $W$ of the image in terms of pixels, 
\begin{equation}
\bigtriangleup_u \leq \frac{W \bigtriangleup_x z_{\min}}{2f}.
\end{equation}
Putting the above constraints together, 
\begin{equation}
\bigtriangleup_u \leq \min \left(\frac{D}{2 K_x f \left(1/z_{\min} - 1/z_{\max}\right)},\frac{W \bigtriangleup_x z_{\min}}{2f}\right).
\end{equation}

It is useful to interpret the required camera sampling rate in terms of the maximum pixel disparity $d_{\max}$ of any scene point between adjacent input views. If we set $z_{\max}=\infty$ to allow scenes with content up to an infinite depth and additionally set $K_x = 1/\left(2\bigtriangleup_x\right)$ to allow spatial frequencies up to the maximum representable frequency, 
\begin{equation}
\frac{\bigtriangleup_u f}{\bigtriangleup_x z_{\min}} = d_{\max} \leq \min\left(D,\frac{W}{2}\right).
\end{equation}
Simply put, the maximum disparity of the closest scene point between adjacent views must be less than $\min(D,W/2)$ pixels. When $D = 1$, this inequality reduces to the Nyquist bound: a maximum of 1 pixel of disparity between views, but in general can support a disparity of $D$ pixels for a $D-$layer MPI.  Note that we are also bounded by the width of the image, and the disparity cannot exceed $W/2$ regardless of the number of layers in the MPI; this is needed for field-of-view overlap for geometry estimation. Finally, note that real scenes must be sampled in 2 dimensions for the camera, leading to a sampling reduction of $D^2$.  If $D=64$ as in most of our examples, this leads to a sampling rate reduction of more than $4000\times$, making view synthesis practical.  

Having established the fundamental theory behind our method, let us turn to the practical algorithm.  The expansion of each view to a local light field using an MPI scene representation is performed using a simple convolutional neural network taking five views as input: the reference view to be expanded and the four nearest neighbors.  Note that this is the only black box use of deep learning in the method and could be replaced by any other MPI construction scheme.  While the paper works with deep learning, it is not fundamental to the contributions of the work, and we therefore do not discuss this part of the algorithm in detail.  {\em It is important to note that the local light field fusion paper is a rare example of work within deep learning that provides a fundamental theoretical analysis and guarantees on the sampling rates for image capture.}

The name {\em local light field fusion} derives from a simple generalization of the above algorithm, where local light fields for each view are blended together to enable more accurate light field reconstruction over the entire volume where cameras capture the scene.  This enables the full generality of light field rendering, with new view paths having unconstrained 3D translation and rotation.  Please refer to the paper for details on computing blending weights for combining local light fields.  We also refer the reader to the original paper in terms of the training procedure and datasets, simply remarking that the amount of data needed is substantially lower than most deep learning methods, and consists largely of synthetic data, with only 24 real scenes for fine-tuning.  

\begin{figure}
\begin{centering}
\includegraphics[width=0.5\columnwidth]{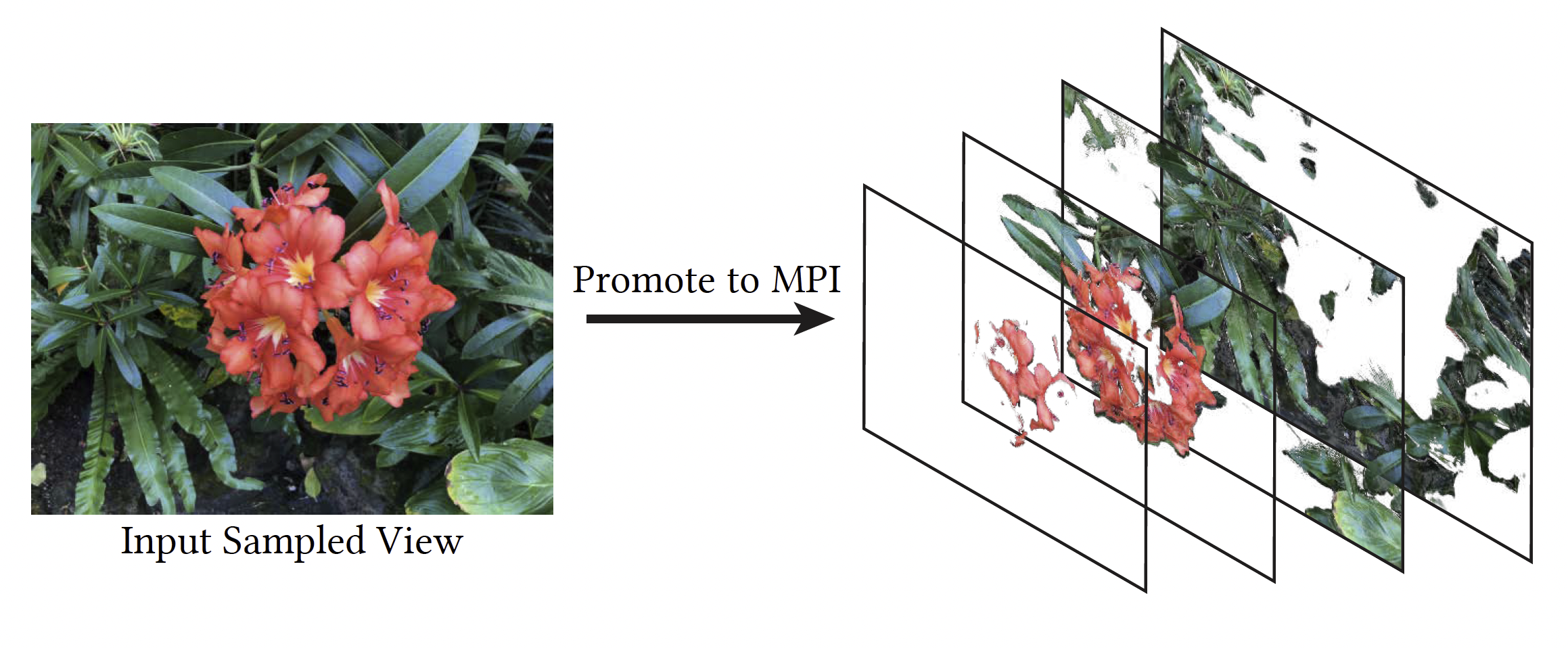} 
\includegraphics[width=0.5\columnwidth]{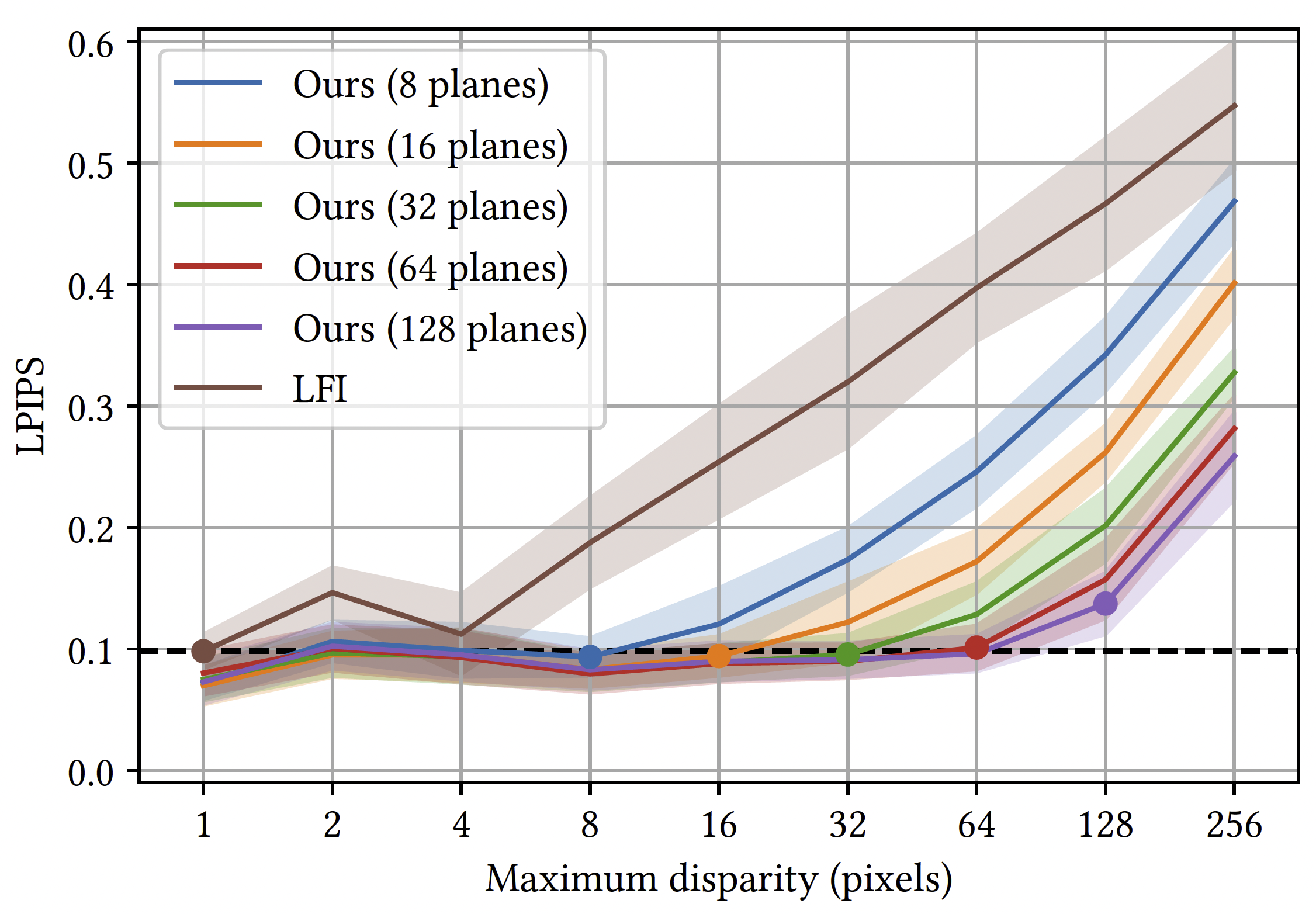} 
\end{centering}
\vspace*{-.1in}
\caption{\small \em Left: The basic idea of lifting an input sampled view to a multiplane image with RGB color and opacity.  Right: Validation of the method and theory, showing that with a $D$ layer (or planes) MPI, we can reconstruct scenes up to a disparity of $D$ pixels, at least until $D=64$, with the same perceptual quality as light field interpolation with Nyquist rate sampling (black dotted line).  Note that sampling is in two dimensions, so we achieve the same results as Nyquist rate view sampling with $64^2=4096\times$ fewer views.  For higher numbers of planes, the overlap between adjacent views decreases and errors increase.  The colored dots indicate the point on each line where the number of planes equals the maximum scene disparity, while the shaded region indicates 1 standard deviation over all 8 test scenes.}
\label{fig:validation}
\end{figure}

\begin{figure}
\begin{centering}
\includegraphics[width=\columnwidth]{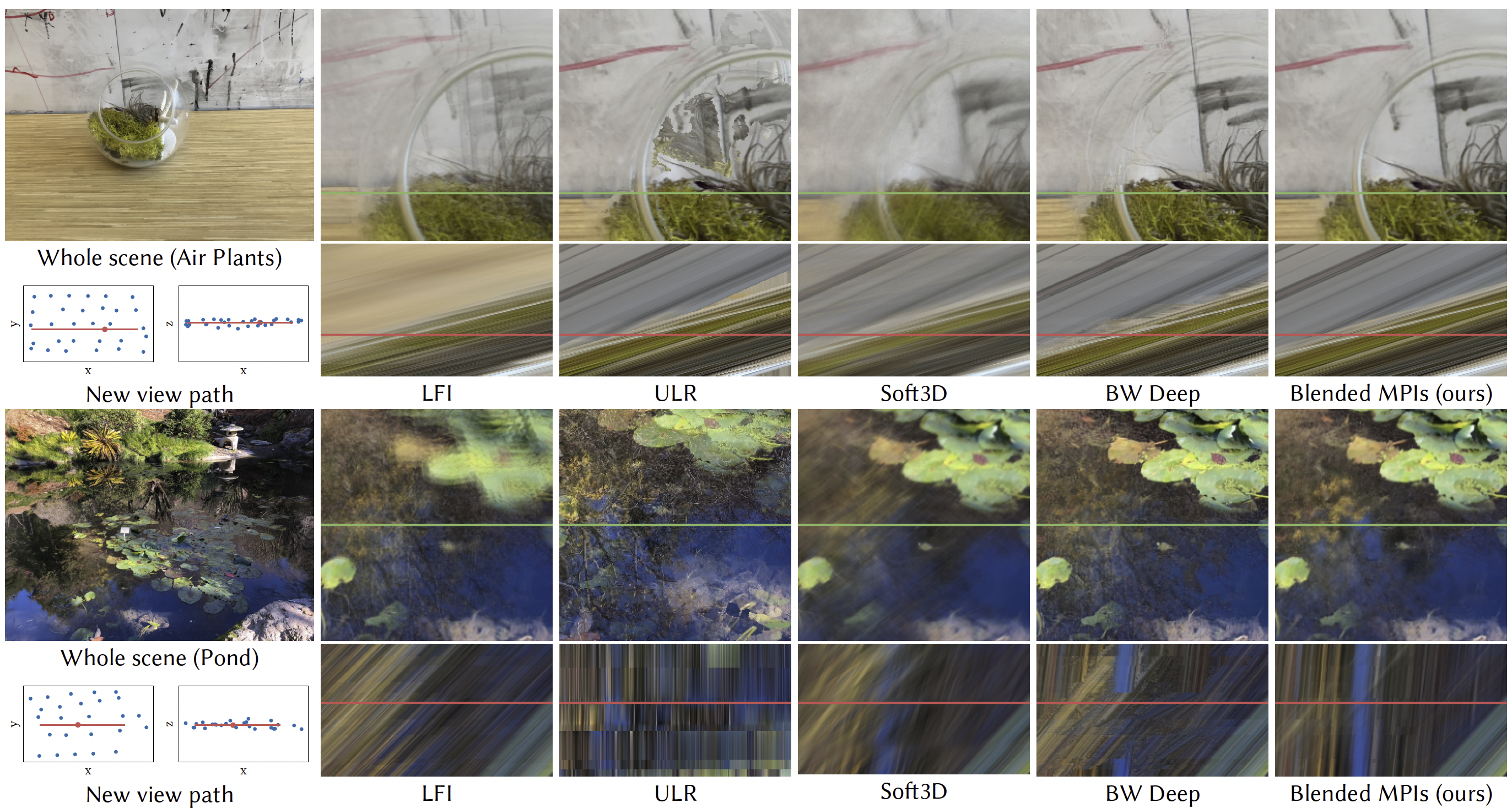} 
\end{centering}
\caption{\small \em Results on two scenes (from Fig. 9 of the original paper~\cite{llff}).  These datasets were captured by a standard cellphone.  We render a sequence of new views and show both a crop from a single rendered output and an epipolar slice of the sequence. We show 2D projections of the input camera poses (blue dots) and new view path (red line) along the z and y axes of the new view camera in the lower left of each row.  Comparison is made to prior methods, showcasing the quality of results from local light field fusion.}
\label{fig:results}
\end{figure}

The core result of the paper is perhaps the validation shown in Fig.~\ref{fig:validation} (right).  This shows that we can render novel views with Nyquist level perceptual quality with up to $d_{\max}=64$ pixels of disparity between input view samples, as long as we match the number of planes in each MPI to the maximum pixel disparity between input views.  This validates the theoretical analysis in the paper, and shows that the simple deep learning method proposed for MPI construction can indeed be practical and consistent with plenoptic sampling theory.  A couple of results on real scenes from the paper are shown in Fig.~\ref{fig:results} and also compared to prior work, showcasing the benefits with a relatively small number of input images, including in scenes with complex occlusions and non-Lambertian effects.  

Ultimately, the practical benefit of local light field fusion is in providing prescriptive scene sampling guidelines.  The paper shows that for a smartphone camera with a $64^\circ$ field of view and an MPI with $64$ planes, one can simply write, 
\begin{equation}
\frac{W}{\sqrt{N}} \leq \frac{80 z_{\min}}{S},
\end{equation}
where $W$ is the image resolution (width in number of pixels), $N$ is the total number of images, $z_{\min}$ is the minimum depth of objects in the scene and $S$ corresponds to the side length of a world space plane that bounds the viewpoints we seek to render.  Once the user has determined the extent of viewpoints they wish to render and thus fixed $S$, and image resolution $W$ is known, we can determine the number of images $N$ based on the minimum depth $z_{\min}$.  The paper discusses further details on asymptotic rendering time and space complexity.  

Finally, the paper demonstrated a smartphone app for iOS, based on these guidelines, using the ARKit framework that also estimates $z_{\min}$.  We use built-in software to track the phone's position and orientation, providing sampling guides that allow the user to space photos evenly at the target disparity.  Once the user has centered the phone so that the RGB axes align with one of the guides, the app automatically captures a photo.  Moreover real-time viewers have been implemented on both desktop and mobile devices.  

\begin{figure}
\begin{centering}
\includegraphics[width=0.38\columnwidth]{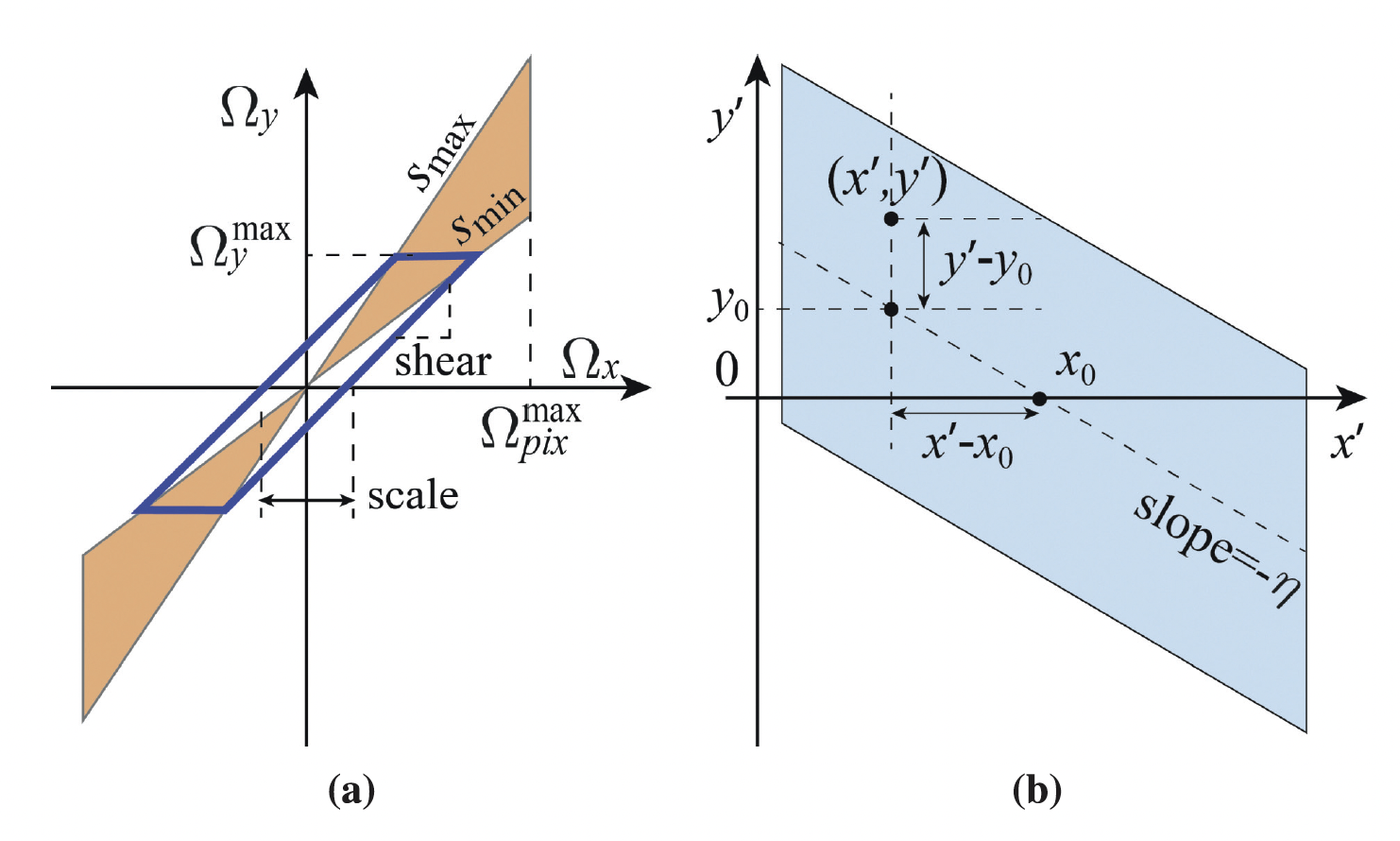} \hskip 0.5in
\includegraphics[width=0.57\columnwidth]{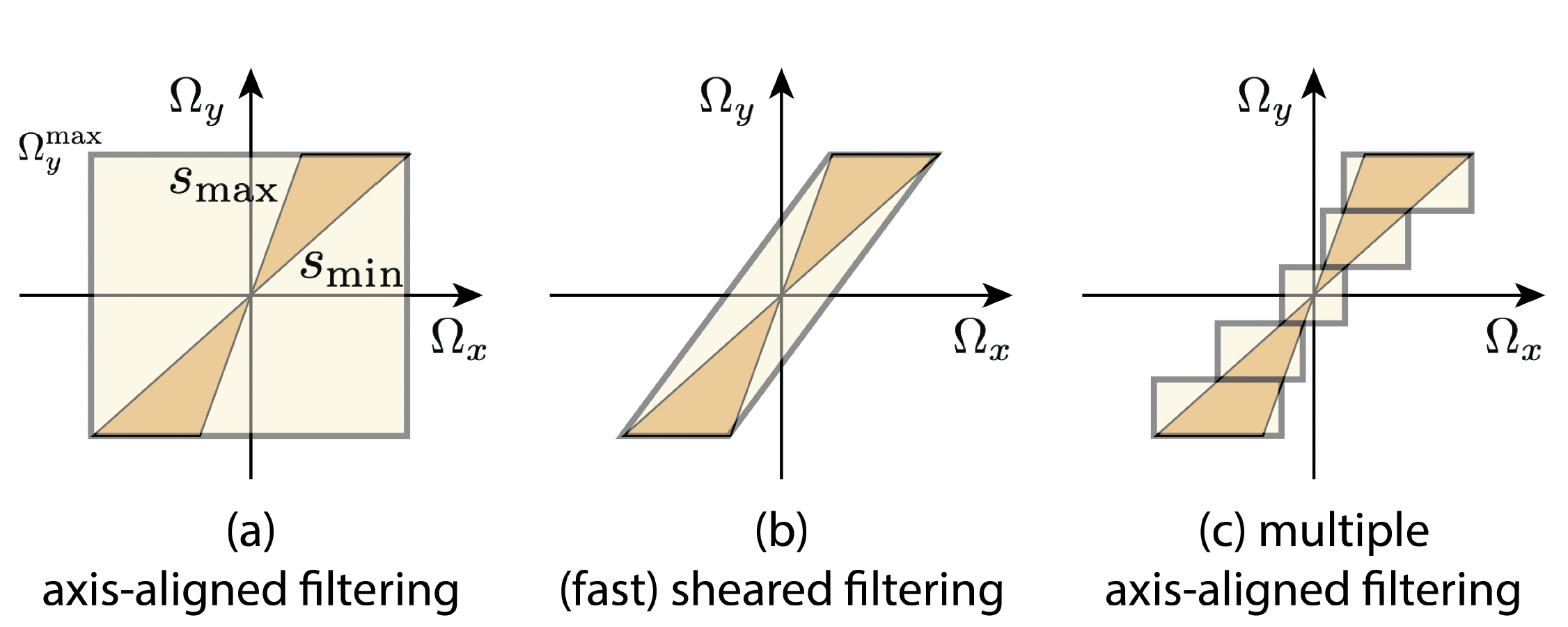} 
\end{centering}
\caption{\small \em Sheared and Multiple Axis-Aligned Filtering for sampling and reconstruction in Monte Carlo Rendering and Denoising.  On the left, we show the sheared Fourier filter, and the corresponding parallelogram filter in the primal domain (from fast 4D sheared filtering~\cite{yan2015fast}).  On the right, we show approximation with multiple axis-aligned filters (from MAAF~\cite{maaf}), and comparison to axis-aligned and sheared filtering.}
\label{fig:filtering}
\end{figure}

\section{Analogy with Sampling for Monte Carlo Rendering and Denoising} 
Having described the basic local light field fusion algorithm, we will now provide some insights and a brief perspective on related work and subsequent efforts.  We note again that this is not intended to be a comprehensive survey, but rather some specific thoughts and comments from the author of this article.  

First, we draw analogies to a somewhat different area of physically-based Monte Carlo rendering, where sampling, reconstruction and denoising are typically used nowadays to create synthetic computer graphics imagery for both real-time applications like games and offline applications like movies.  Since few researchers have worked on both Monte Carlo rendering and view synthesis, these connections are not usually well appreciated, and so we seek to briefly highlight them in this article.  Moreover, this will provide background for some of our later thoughts in terms of sampling analysis of newer methods.  For a somewhat dated survey on Monte Carlo sampling and reconstruction, readers are referred to~\cite{ZwickerSTAR}.  

My (Ravi Ramamoorthi's) group has been working on a sample-and-reconstruct framework for low sample count Monte Carlo rendering (now known as Monte Carlo denoising), starting with our work on motion blur in 2009~\cite{Egan}.  In that work, we showed how the same plenoptic sampling theory explored subsequently for local light field fusion can be used to analyze the space-time description of an object moving under motion blur (including shadows and reflections).  This then leads to an analysis of sampling rates, and appropriate sheared reconstruction filters, just like in view synthesis and image-based rendering.  My group had a subsequent series of works where we showed that the double-wedge frequency spectrum applied to most other visual effects, such as space-angle light fields for soft shadows~\cite{Eganshadows} or directional occlusion~\cite{Eganamb}, building on pioneering work on a frequency analysis of light transport by Durand and collaborators~\cite{Fredo}.  Figure~\ref{fig:filtering} is taken from some of the later papers in the series on fast sheared filtering~\cite{yan2015fast} and multiple axis-aligned filtering~\cite{maaf}.  Figure~\ref{fig:filtering}(a) explains the basic idea of a double wedge frequency domain spectrum reconstructed with a sheared parallelogram filter.  This is also a sheared filter in the primal domain (b) where one shares information across multiple pixels, while projecting samples using the motion (for motion blur, or equivalent for other effects).  An alternative is to pack the filter tightly with multiple axis-aligned boxes that can be easier to evaluate and more compact (Fig~\ref{fig:filtering}(c,d,e)).  

As in view synthesis, these insights on the frequency domain filter enable a substantially sparser sampling rate, often just a few samples per pixel, rather than the hundreds or thousands needed in conventional path tracing, followed by filtering or reconstruction using the correct filter, a process currently often referred to as Monte Carlo denoising.  

The subsequent development of Monte Carlo denoising also has many parallels to view synthesis.  The hand-crafted frequency analysis for specific effects has today largely been replaced with general deep learning approaches, first presented at SIGGRAPH 2017~\cite{learn1,learn2} which take in guide buffers like position, normals, depth, textures to enable higher-quality reconstruction than simple image denoising.  These methods often predict the kernel for filtering, in effect automating the Fourier analysis approach that analyzes the signal, then develops the appropriate sheared or axis-aligned kernel with suitable bandwidth.  
It is interesting that these methods no longer provide sampling guarantees in terms of the required sampling rate, as in the frequency-space approaches.\footnote{For a recent approach that does provide a stopping criterion, see~\cite{Firmino}.}  However, they have enabled accurate reconstruction of Monte Carlo effects, often with only one sample per pixel.  This in turn has led to a revolution in the use of physically-based rendering and low sample-count path tracing in production rendering for movies, where almost every pixel in computer-generated animations and films is now path-traced.  Increasingly, physically-based rendering and path tracing is also used in interactive applications like video games, with the first fully path-traced games hitting the market.  These developments have of course been aided by advances in GPU architectures, including support for real-time raytracing, and denoisers and supersampling techniques in software and hardware from major vendors.\footnote{For example, NVIDIA's RTX architecture for real-time raytracing and DLSS for deep learning super sampling, which also enables reconstruction and denoising.}  As will be discussed next, subsequent developments in view synthesis have proceeded along similar lines, with the expectation of a similarly outsize impact.  

\section{Perspective on Further Advances in View Synthesis and Challenges}
We now very briefly discuss further work on view synthesis.  The subsequent year after the local light field fusion paper, we developed the neural radiance field method for view synthesis (NeRF)~\cite{nerf,nerfcacm}, which has become the method of choice for view synthesis and a variety of related topics well beyond computer graphics.  A number of new repreentations have also been introduced, of which perhaps the best known are instant neural graphics primitives~\cite{instantNGP} and gaussian splatted radiance fields~\cite{Kerbl}.  Note that the NeRF paper is now approaching 10,000 citations and this is by no means an exhaustive list of the work in the area.  NeRF was recognized by a Frontiers of Science Award in 2023, and we encourage readers to read my perspective on that~\cite{ravinerf}, which also has links to surveys.  

As noted in joint first-author Pratul Srinivasan's ACM award-winning dissertation, NeRFs can be seen as a natural evolution in the 3D scene representation employed for view synthesis.  Indeed, Pratul started out by predicting 4D light fields directly (notably in his single-image work~\cite{PratulICCV}).  However, this has limitations, so the next step is to search for a persistent 3D scene representation, for which multiplane images (MPI), as used in local light field fusion, were an ideal candidate.  In retrospect, multiplane images or MPIs can be seen as a discrete volumetric radiance field, sampled at a number of depth layers along the z-direction while using a pixel sampling along x and y.  The core idea however is in terms of a volumetric rather than surface representation, in order to handle ambiguity and errors in surface shape, using the alpha or opacity channel to effectively blend different layers.  

The MPI representation does have limitations in terms of being discrete, in effect with many similarities to a discrete voxel grid.  Neural radiance fields~\cite{nerf} instead demonstrate a continuous volume, where a simple multi-layered perceptron takes in the spatial coordinates and angular direction and outputs color and the volume density (technically, the extinction coefficient), which can be related to the opacity for alpha-blending.  This provides a more compact representation, and NeRFs store the entire volume within 5MB, in terms of the weights of the MLP network.  Subsequent work has explored hybrid grid-MLP methods, as in instant NGP~\cite{instantNGP} or simply using a combination of Gaussians as a Gaussian-splatted radiance field~\cite{Kerbl}.  The last approach does not even require any MLPs or other machine learning representations.  It is also possible to view Gaussian splatting within the general framework of raytracing volumes~\cite{gaussianrt}. Again, we can only touch very briefly on these exciting developments and refer readers to my earlier article and linked surveys for more details, and for a discussion of a plethora of additional representations that have been proposed in the past few years.   

The impact of these developments on view synthesis has been undeniable, with startups such as Luma AI enabling users to take a few images on their phone and create 3D models, and Google using these methods within both Streeview and shopping apps.  Lay users are now able to create NeRFs, and the New York times has blogged about them for portrait capture, as just a snapshot of the impact and exciting applications.  Perhaps most interesting, these representations form a bridge between 2D images and 3D models that can be exploited by the new wave of generative AI techniques, enabling generative AI methods trained on large-scale collections of 2D images to scale up to generating 3D data.  The impact is only expected to increase over time, and numerous other applications like dynamic scenes, acquisition of full light transport for relighting, and new augmented and virtual reality experiences have also been explored.  

Since this article is devoted to the local light field fusion method, we will however focus the remainder of this paper on the question of sampling rates.  Indeed, the goal of local light field fusion was to reduce the Nyquist sampling rate by orders of magnitude in a principled way to enable sparse capture.  This is akin to Monte Carlo rendering discussed earlier, where light field signal processing theory and Monte Carlo denoising have enabled substantially lower sample counts.  Like in Monte Carlo rendering, the first wave of methods based on principled light field theory achieved sampling rates of a few tens of images (samples per pixel for Monte Carlo).  This was in itself a remarkable advance.  However, the next generation of algorithms based on deep learning showed an even more dramatic reduction in sample counts, down to one sample per pixel in Monte Carlo rendering, and we are taking a similar trajectory with corresponding impact in view synthesis.  

To provide just some examples in recent work, our recent paper enables extremely sparse sampling rates, often just 3-6 images, and does not require the initial step of estimating camera pose (Jiang et al.~\cite{Jiangsig}).  Even for the single-image case, remarkable advances has been made in the past year.  My group had a few early papers~\cite{PratulICCV,nerfdiff,kaienwacv}, and current methods~\cite{Vondrick,zero45plusplus,cat3D} have demonstrated remarkable fidelity from only a single image.  This is in many ways the holy grail, enabling one to take legacy 2D photo collections and turn them into immersive 3D experiences, or to use 2D image generators in generative AI, and create 3D versions for free.  A number of startups and established companies are exploring all of these directions, in addition to academic research.  The analogy with Monte Carlo rendering is also clear, in that just as those methods have pushed to one sample per pixel rendering as the default, at least for real-time applications, view synthesis has made remarkable progress in reducing the number of views needed by several orders of magnitude, pushing towards a single input image in the limit.  

We close this section and article by raising an open challenge.  Local Light Field Fusion was recognized with the Frontiers of Science Award for providing rigorous prescriptive view sampling guidelines based on frequency analysis and suitable Nyquist limits.  Just as in Monte Carlo rendering, we have seen an evolution away from explicit frequency-space analysis to more learning-based approaches that can provide dramatic results with one sample per pixel or a single view.  However, these newer methods, starting with NeRF and going all the way to present techniques, {\em provide no sampling guarantees nor prescriptive guidelines for where views should be taken.}  This is the same situation in Monte Carlo rendering where for the most part, current deep learning denoising methods provide no analysis of required sampling rates or guarantees of convergence.  As such, while we have made substantial experimental progress, we have lost the theoretical understanding and guarantees of local light field fusion.  {\em This remains an open challenge to the community, in terms of quantifying the required sampling rate or sampling-error curves in newer volumetric radiance field algorithms for view synthesis, and providing guarantees on sampling.}

\section*{Conclusion}

The ability to take a few photographs and capture the appearance of a real scene, to then be able to re-render it seamlessly from other viewpoints is a key challenge in computer graphics, computer vision and virtual reality, referred to as image-based rendering or view synthesis.  We have been fortunate to receive back-to-back frontiers of science awards for our papers on local light field fusion and neural radiance fields to address this problem.  The current article pertains to the SIGGRAPH 2019 paper on sampling for view synthesis with local light field fusion, where the key contribution is actually a frequency domain analysis of the light field or plenoptic function, which enables prescriptive guidelines regarding how many images to take and where to sample views.  In particular, we show that by predicting a 64-layer multiplane image, one can reduce the number of views needed by $64^2$ or $4096\times$, which enables view synthesis from a sparse set of images, and makes the method practical with rigorous sampling gurantees.  This paper also represents one of the only works with a deep learning component that provides formal theoretical analysis.

Subsequent work on neural radiance fields and extensions has generalized the discrete volumetric representation of the multiplane image to continuous volumetric representations that have provided the highest visual quality for view synthesis, even enabling a number of novel applications in domains well outside computer graphics, and connecting with advances in modern artificial intelligence.  Besides practical applications, a slew of current methods enable very sparse view synthesis, with excellent results often available from only a single input image, with the potential to take legacy 2D photographs and promote them to full 3D immersive experiences.  

In many respects, this progression parallels the development of methods for Monte Carlo image denoising that started with similar theoretical foundations based on a frequency analysis of light transport, followed by deep learning approaches that generalized the earlier methods and drove sample counts down dramatically to one sample per pixel.  In both cases, the tremendous practical progress has overtaken theory, and there are unfortunately very limited or no theoretical foundations on sampling rates and rigorous sampling guarantees for current methods in either view synthesis or Monte Carlo rendering.  At one level, this is unavoidable; once one is working in the limit of a single input image or one sample per pixel, it is unclear if there is a meaningful sampling theory.  However, in both cases, image quality does improve as more samples are taken, and a theoretical or even empirical analysis of the image quality versus number of samples tradeoff, comparable to the original plenoptic sampling paper, would be very welcome.  

\section*{Acknowledgements: } 
First, I of course acknowledge my co-authors on the paper, my then Ph.D. student Pratul Srinivasan, Ben Mildenhall, former postdoc Nima Kalantari, Rodrigo Ortiz-Cayon, Ren Ng and Abhishek Kar.   I especially thank Rodrigo for traveling to Beijing with me for the Frontiers of Science Award and to Nima for providing a sounding board.  I also wish to acknowledge all of the students and postdocs who contributed and continue to do so to light field and view synthesis/appearance projects, including Michael Tao, Ting-Chun Wang, Matt Tancik and many others.  The close relation of local light field fusion with Monte Carlo rendering and seeing the links between them would not have been possible without the work of Nima, as well as my Ph.D. students Kevin Egan, Soham Mehta, Lingqi Yan and many others.  I express my thanks to all of my colleagues and students within the UC San Diego Center for Visual Computing.  I want to thank our funding agencies all through the years; the original paper acknowledges NSF grants 1617234 and 1617794, ONR grant N000141712687, Google Research Awards as well as fellowships (Hertz to Ben Mildenhall, NSF to Pratul Srinivasan and Sloan to Ren Ng).  I also wish to acknowldege the support of the Ronald L. Graham Chair, and our industrial partners over the years, including but not limited to, Google, Adobe, Samsung, Qualcomm, Sony, Draper and Fyusion.  ONR program manager Behzad Kamgar-Parsi has funded our research for close to two decades, including the grants key to our light field and view synthesis projects.  The original paper also acknowledges Julius Santiago, Milos Vlaski, Endre Ajandi and Christopher Schnese for producing the technical video (an excerpt of which was shown at the evening of computer science at ICBS 24), and Alex Trevor for developing the augmented reality application.  

We thank the organizers of the International Congress on Basic Science for envisioning a unique conference, recognition of view synthesis with two frontiers of science awards, and for their exceptional dedication to basic science and the extraordinary hospitality shown at the conference.  Perhaps the greatest thanks should go to the thousands of researchers who have worked on image-based rendering and view synthesis for at least the past three decades, bringing the field to its current exciting juncture, even as we look forward to even greater advances in the years to come.   
\bibliography{combined}
\bibliographystyle{plain}

\end{document}